\documentclass[aps,twocolumn]{revtex4}
\usepackage[utf8]{inputenc}
 \usepackage{amsmath,amssymb}
 \usepackage{graphicx}
 \usepackage{float}
\usepackage{color}
\usepackage[colorlinks=true,citecolor=blue,linkcolor=blue,urlcolor=black]{hyperref}
\begin{document}                                                                                

\title{Constraints on R\'{e}nyi Entropy through Primordial Big-Bang Nucleosynthesis and Baryogenesis}
\author{Ahmad Sheykhi\footnote{asheykhi@shirazu.ac.ir} and Ava Shahbazi Sooraki}
\address{Department of Physics, College of
Science, Shiraz University, Shiraz 71454, Iran\\
Biruni Observatory, College of Science, Shiraz University, Shiraz
71454, Iran}

 \begin{abstract}
The R\'{e}nyi entropy, a one-parameter generalization of
Boltzmann-Gibbs entropy, offers a promising framework for probing
quantum gravitational effects in cosmology. By modifying the
entropy-area relation of the apparent horizon, R\'{e}nyi entropy
can alter the expansion dynamics of the early universe, with
potential implications for Big-Bang Nucleosynthesis (BBN). In this
work, we derive the modified Friedmann equations within the
R\'{e}nyi entropy paradigm and investigate their impact on the
primordial abundances of light elements Deuterium ($D$), Helium-4
($_{}^{4}\textit{He}$), Lithium-7 ($_{}^{7}\textit{Li}$) and
baryogenesis. Using observational constraints from Planck,
primordial abundance data and observational data on baryogenesis,
we put stringent bounds on the R\'{e}nyi parameter. Furthermore,
we explore whether R\'{e}nyi entropy corrections could mitigate
the long-standing Lithium discrepancy. This study provides the
first systematic constraints on R\'{e}nyi cosmology from BBN and
highlights the role of nonextensive thermodynamics in
early-universe physics. Our analysis shows that the obtained
ranges for the R\'{e}nyi parameter $\lambda$ exhibit a overlap for
Helium-4 and Deuterium, but this overlapping region-despite a
small discrepancy-is inconsistent with the range obtained from
Lithium-7.  This small mismatch between the ranges raises the
possibility of alleviating the \textit{Lithium Problem} in the
modified cosmology scenarios. Furthermore, we present the
relationship between the cosmic time and temperature within the
framework of R\'{e}nyi cosmology. We observe that an increase in
the R\'{e}nyi parameter increases the temperature of the early
Universe.

\end{abstract}
 \maketitle
\section{Introduction\label{Intro}}
Black hole thermodynamics has provided compelling evidence that
gravitational dynamics may emerge from fundamental thermodynamic
principles. The close relation between horizon geometry and
thermodynamic quantities such as entropy and temperature has
motivated the conjecture that gravitational field equations act as
equations of state. A seminal result in this direction was
obtained by Jacobson, who derived Einstein's equations from the
Clausius relation $\delta Q = T\,\delta S$ applied to local
Rindler horizons \cite{Jac}. This approach was later extended to a
variety of modified theories of gravity, including $f(R)$ gravity
\cite{Elin}, Gauss-Bonnet gravity, scalar-tensor theories, and the
general Lovelock class \cite{Pad1,Pad2,Cai1,Pad3}.

In cosmology, this correspondence becomes particularly
transparent. Within Friedmann-Robertson-Walker (FRW) spacetime,
the Friedmann equations can be written in the form of the first
law of thermodynamics at the apparent horizon $ dE = T_h\, dS_h +
W dV $ and, conversely, gravitational dynamics can be recovered
from this thermodynamic identity \cite{Frolov, Calcagni,Cai2,
Sheykhi1,Sheykhi2,Sheykhi3,Sheykhi4,wang1,wang2,CaiKim}. In this
framework, the entropy of the horizon is obtained by adopting the
black hole entropy of the underlying gravity theory and replacing
the black hole horizon radius $r_{+}$ with the apparent horizon
radius $\tilde r_{A}$. The standard Bekenstein-Hawking area law $
S_h = {A}/{4G} $ provides the simplest description, but it is well
known that this relation receives important corrections.

Quantum gravitational effects generate logarithmic and power-law
corrections to the entropy-area law. Logarithmic corrections arise
from thermal and quantum fluctuations, particularly in loop
quantum gravity \cite{Das,Ashtekar,Zhang,Banerjee,SheyLog}, while
power-law corrections naturally appear due to the entanglement of
quantum fields across the horizon \cite{Das2,Radicella,SheyPL}.
Moreover, gravitational systems are characterized by long-range
interactions and often divergent partition functions, which
motivate the replacement of the additive Boltzmann-Gibbs entropy
with generalized, nonadditive entropy formalisms
\cite{Wilk,Gibbs,RNunes,Tsallis1}. These ideas are closely related
to holographic and emergent gravity perspectives. Verlinde's
proposal of entropic gravity showed that gravity can be
interpreted as an entropic force arising from changes in
information associated with holographic screens, based on the
holographic principle and the equipartition law \cite{Ver}.
Padmanabhan further developed this viewpoint by proposing that the
expansion of the universe is driven by the emergence of cosmic
space, governed by the difference between the degrees of freedom
on the boundary and in the bulk \cite{PadEm}. By equating this
difference to the change in effective volume, one recovers the
Friedmann equations in a thermodynamically transparent way
\cite{CaiEm,Yang,Sheyem}.

More recently, it has been recognized that generalized entropies
motivated by quantum gravity and nonextensive statistical
mechanics can significantly modify the cosmological dynamics
\cite{Odin1, Odin2, Odin3, Odin4}. In particular, Tsallis entropy
\cite{Tsallis1}, which is intrinsically nonadditive and is based
on the principles of nonextensive statistical mechanics, together
with Barrow entropy \cite{Barrow0} and Kaniadakis entropy
\cite{Kan1,Kan2}, introduce deformation parameters that quantify
deviations from the Shannon-Gibbs law and lead to modified
Friedmann equations. Within the Tsallis framework, the horizon
entropy acquires a nonadditive, power-law deformation that
directly alters the Friedmann equations \cite{Sheykhi1}. Barrow
entropy, inspired by fractal deformations of horizon geometry
associated with spacetime foam, introduces an effective dimension
parameter that modifies the cosmological field equations
\cite{Emm2, Sheykhi2, SheB2}. Kaniadakis statistics, based on a
relativistic deformation of phase-space structure, likewise leads
to corrections in the expansion history of the universe
\cite{Lym,Her,Dre,ShKa1,ShKa2,Salehi}. These potential imprints
make primordial nucleosynthesis a powerful laboratory for testing
generalized entropic cosmologies. Indeed, modifications to the
expansion rate $H(T)$ during BBN can directly alter the freeze-out
of nuclear reactions, changing the predicted abundances of light
elements such as Deuterium, Helium-4, and Lithium-7, and modify
the relation between cosmic time and temperature $t(T)$ in the
early universe.

The standard cosmological model provides a remarkably successful
description of the universe's evolution, with BBN serving as one
of its earliest and most robust observational pillars. Occurring
during the first few minutes of cosmic history $(T \sim 0.1-1$
MeV), BBN predicts the primordial abundances of light elements,
Deuterium ($D$), Helium-4 ($_{}^{4}\textit{He}$), and Lithium-7
($_{}^{7}\textit{Li}$) with striking agreement with observations
for $D$ and $_{}^{4}\textit{He}$ \cite{RH1, RH2, Planck}. However,
a persistent discrepancy, known as the cosmological\textit{
Lithium problem}, remains unresolved, the predicted
$_{}^{7}\textit{Li}$ abundance exceeds observations by a factor of
2-3 \cite{Coc}. This anomaly suggests either unknown systematic
errors in astrophysical measurements or new physics beyond the
standard BBN framework, such as modified gravitational theories or
nonextensive thermodynamics \cite{Barrow,Luciv}.

This connection between modified entropies and BBN has already
been exploited to constrain several generalized entropy models.
For instance, bounds on the Tsallis parameter from primordial
light elements were explored in \cite{Anish}, while cosmological
consequences on baryogenesis and ${_{}^{7}\textrm{Li}}$-abundance
in the context of (dual) Kaniadakis-modified cosmology were
studied in \cite{Luciv,Ava2}. The latter analysis revealed that
the allowed values for the dual Kaniadakis parameter fall within
the extremely narrow range $-0.8 \times 10^{-78} \lesssim
\tilde{K^*} \lesssim 0.8 \times 10^{-78}$, indicating minimal
deviations from standard thermodynamics \cite{Ava2}. Similarly,
constraints on the Barrow exponent $\delta$ from BBN data suggest
$\delta \simeq 0.01$ to avoid spoiling the nucleosynthesis era
\cite{Barrow,Ava1}. Despite these investigations into Tsallis,
Barrow, and Kaniadakis entropies, a systematic analysis of
R\'{e}nyi entropy in the context of BBN remains absent.

The R\'{e}nyi entropy \cite{Reny}, a one-parameter generalization
of the Boltzmann-Gibbs entropy, offers a compelling avenue to
explore deviations from standard cosmology. Originally introduced
in information theory, R\'{e}nyi entropy  has found applications
in black hole thermodynamics \cite{Kom1,Moradpour1}. When applied
to the apparent horizon of the universe, R\'{e}nyi entropy
modifies the gravitational field equations via the
thermodynamics-gravity conjecture \cite{Moradpour2,Nae,faz},
potentially altering the expansion rate $H(T)$ during BBN and thus
the primordial abundances of light elements.

In this work, we address the question, how do R\'{e}nyi entropy
corrections influence BBN predictions through modification of
cosmological field equations? Our focus is to apply the
observational constraints from primordial nucleosynthesis and
baryogenesis to R\'{e}nyi cosmology and to impose observational
bounds on the R\'{e}nyi parameter. For this purpose, we first
derive the modified Friedmann equations from R\'{e}nyi entropy and
compute the resulting changes to the time-temperature relation,
neutron freeze-out, and light-element abundances. The implications
of the resulting $\lambda$-dependent Friedmann equations are
examined in relation to baryogenesis and the primordial abundance
of the light elements. By comparing these predictions with
observational constraints on $D/H$, $_{}^{4}\textit{He}$,
$_{}^{7}\textit{Li}$ and the baryogenesis observational data, we
place stringent bounds on the R\'{e}nyi parameter.  Our analysis
also reveals whether R\'{e}nyi entropy can alleviate the
\textit{Lithium problem} while remaining consistent with other
elemental abundances. This study provides the first systematic
test of R\'{e}nyi cosmology against BBN data, offering insights
into the role of nonextensive thermodynamics in the early
universe.

The outline of this paper is as follows. In the next section, we
show how Newton's law of gravity gets modified through
R\'{e}nyi-corrected entropy. Interestingly enough, we show that
the theoretical origin of the Modified Newtonian dynamics (MOND)
can be understood via modified R\'{e}nyi entropy. In the
relativistic regime, we obtained the modified Friedmann equation
through thermodynamics-gravity conjecture in section III. In
section IV, we constrain the free parameter of R\'{e}nyi cosmology
using primordial BBN and comparing the results with observational
data. We also examine the Lithium problem in the context of
modified R\'{e}nyi cosmology in this section. Section V is devoted
to study baryogenesis and constraining R\'{e}nyi parameter by
requiring consistency between theoretical predictions and
baryogenesis observational data. In section VI, we investigate how
the relation between time and temperature alter due to change in
the expansion rate, $H(t)$. We finish our paper with closing
remarks in section VII.
\section{MOND theory from R\'{e}nyi entropy \label{Ren}}
The Modified Newtonian dynamics (MOND) suggested by Milgrom to
explain the flat rotation curves of the spiral galaxies
\cite{Milgrom1}. According to the MOND theory, the Newton's second
law get modified for the large scales as
\begin{equation}\label{F0}
F=m  \mu \left(\frac{a}{a_0}\right)a,
\end{equation}
where $a$ stands for the usual kinematical acceleration, which is
taken as $a=v^{2}/R$, and $a_{0}=1.2\pm 0.27  \times 10^{-10}$
$m/s^{2}$ is a constant \cite{Begeman}. Here $\mu(x)$ is a real
function satisfies the following boundary conditions,
\begin{equation}\label{mu0}
\mu(x)\approx\left\{
  \begin{array}{ll}
  $$1$$  \quad \quad {\rm for} \quad\quad {x}\gg 1, &  \\
  $$x$$  \quad \quad {\rm for}  \quad\quad {x}\ll 1.
    \end{array}
\right.
\end{equation}
At large distance, at the galaxy outskirt, the kinematical
acceleration `$a$' is extremely small, smaller than $10^{-10}$
$m/s^{2}$ , i.e., $a\ll a_{0}$, hence the function $\mu
(\frac{a}{a_{0}})=\frac{a}{a_{0}}$. Mathematically, for a galaxy
with mass $M$ and a star (particle) with mass $m$, the Newton's
law of gravity get modified as
\begin{equation}\label{v}
F=m \frac{a^2}{a_{0}}=\frac{GMm}{R^2}, \Rightarrow  v= (GM
a_{0})^{1/4}\approx cte.
\end{equation}
This implies that the velocity of a star, on circular orbit from
the galaxy-center is constant and does not depend on the distance;
the rotation curve is flat, as observed.

Here, we are going to show that the theoretical origin of the MOND
theory can be completely understood using the modified R\'{e}nyi
entropy. R\'{e}nyi entropy, introduced by the Hungarian
mathematician Alfred R\'{e}nyi in $1961$ \cite{Reny}, is a
generalized measure of uncertainty that extends the classical
Shannon entropy to a broader family of entropy functions.
Originally developed in the context of information theory,
R\'{e}nyi entropy has since found applications in statistical
mechanics, quantum physics, black hole thermodynamics, and complex
systems. Unlike the standard Boltzmann-Gibbs entropy, which is
additive, R\'{e}nyi entropy belongs to the class of nonextensive
entropies, making it particularly useful in systems with
long-range correlations, fractal structures, or quantum
gravitational effects.

For a discrete probability distribution $P = (p_1, p_2,... ,p_n)$,
with $\sum_{i}p_{i}=1$, the R\'{e}nyi entropy of order $\alpha$ is
\cite{Reny}
\begin{equation}\label{Reny}
S_{\alpha}(P)=\frac{1}{1-\alpha}\ln \left(\sum_{i=1}^n
{p_{i}^{\alpha}}\right),
\end{equation}
where $\alpha$ is a scaling parameter, often called the order of
the entropy. When $\alpha\rightarrow1$, R\'{e}nyi entropy reduces
to the Shannon entropy
\begin{equation}\label{Shannon}
S_{1}(P)=- \sum_{i=1}^n {p_{i}\ln {p_{i}}}.
\end{equation}
In the context of black hole physics, the R\'{e}nyi entropy
associated with the black hole horizon is given by
\cite{Kom1,Moradpour1}
\begin{equation}\label{SR1}
S_{h}=\frac{1}{\lambda}\ln(1+\lambda S_{BH}),
\end{equation}
where $S_{BH}=A/4G$ is Hawking-Bekenstein entropy associated with
the horizon, and $\lambda$ is called the R\'{e}nyi parameter. When
$\lambda\rightarrow 0$, one recovers the area law of black hole
entropy. The use of Rényi entropy in this gravitational context
follows established approaches to generalized horizon
thermodynamics \cite{Moradpour1, Kom1}. Here, the microcanonical
counting of horizon states remains unchanged; instead, the
thermodynamic ensemble describing the horizon is modified. Rényi
entropy can be viewed as the von Neumann entropy of a deformed
canonical ensemble. Applying the thermodynamics-gravity
correspondence with this deformed ensemble naturally induces
corrections to the gravitational field equations. Thus, Rényi
entropy offers a phenomenological framework through which
nonextensive thermodynamic effects, potentially arising from
quantum gravitational physics, can be explored within classical
cosmological dynamics.

 The nonextensive parameter $\lambda$ plays the role of
the mapping parameter from non-additive Tsallis entropy to the
additive R\'{e}nyi case. Expanding for small value of $\lambda$,
we find
\begin{equation}\label{SR2}
S_{h}=S_{BH}-\frac{\lambda}{2}S_{BH}^2+\frac{\lambda^2}{3}S_{BH}^3-{\cal{O}}(
\lambda^3 S_{BH}^4).
\end{equation}
We consider a system that its boundary is not infinitely extended
and forms a closed surface with spherical geometry. We can take
the boundary as a storage device for information, i.e. a
holographic screen. We also assume at the center of the
holographic screen there is a mass $M$ and at distance $R$, mass
$m$ is located near the screen. Using the entropic force scenario
\cite{Ver}, and taking into account a general form for the entropy
associated with the holographic screen, we can write down the
Newton's law of gravitation as  ($k_B=\hbar=c=1$) \cite{sheyECFE}
\begin{eqnarray}\label{F1}
F&=&\frac{GMm}{R^2}\times 4G  \frac{ dS_h}{dA}\mid _{A=4\pi
R^2}\nonumber\\&&=4 G m  a \times \frac{ dS_h}{dA}\mid _{A=4\pi
R^2},
\end{eqnarray}
where $a=GM/R^2$ is the acceleration of a particle with mass $m$
which rotates at the distance $R$ around the central mass $M$.

Inserting entropy expression (\ref{SR1}) in Eq. (\ref{F1}), we
find
\begin{eqnarray}\label{F2}
F&=&\frac{GMm}{R^2}\times \left(1+\frac{\lambda \pi
R^2}{G}\right)^{-1}.
\end{eqnarray}
This is the modified Newton's law of gravitation inspired by the
R\'{e}nyi entropy. Now we verify that this modified Newton's law
of gravitation can indeed address the origin of the MOND theory
from thermodynamics arguments. Equating Eq. (\ref{F0}) with Eq.
(\ref{F2}), we find
\begin{eqnarray}\label{mu1}
\mu \left(\frac{a}{a_0}\right)=\frac{1}{1+\frac{a_0}{a}},
\end{eqnarray}
where $a_0=\lambda \pi M$ and $a=GM/R^2$. If we define as usual
$x=a/a_0$, we have
\begin{eqnarray}\label{mu1}
\mu \left(x\right)=\frac{x}{1+x},
\end{eqnarray}
which follow the asymptotic behaviour given in Eq. (\ref{mu0}). In
this way, we address the theoretical origin of the MOND theory by
considering the entropy associated with the horizon in the form of
R\'{e}nyi entropy. In the next sections, we explore the
cosmological consequences of the R\'{e}nyi entropy.
\section{Modified Friedmann equations inspired by R\'{e}nyi entropy \label{Fried}}
Let us start with a spatially homogeneous and isotropic spacetime.
The metric of such spacetime is given by the FRW geometry,
\begin{equation}
ds^2={h}_{\mu \nu}dx^{\mu} dx^{\nu}+R^2(d\theta^2+\sin^2\theta
d\phi^2),
\end{equation}
where $R=a(t)r$, $x^0=t, x^1=r$, and $h_{\mu \nu}$=diag $(-1,
a^2/(1-kr^2))$ represents the two dimensional subspace. The
parameter $k$ denotes the spatial curvature of the Universe with
$k = -1,0, 1$,  corresponds to open, flat, and closed Universes,
respectively. The radius of the apparent horizon, which is a
suitable horizon from thermodynamic viewpoint, is given by
\cite{Hay1,Hay2,Bak}
\begin{equation}
\label{radius}
 R=\frac{1}{\sqrt{H^2+k/a^2}},
\end{equation}
where $H=\dot{a}/a$ is the Hubble parameter and dot stands for the
derivative with respect to time. The apparent horizon is a
suitable boundary from thermodynamic arguments \cite{Cai2}. The
temperature associated with the apparent horizon is defined, by
using the surface gravity, as \cite{Hay1,Hay2,Bak}
\begin{equation}\label{T}
T_h=\frac{\kappa}{2\pi}=-\frac{1}{2 \pi R}\left(1-\frac{\dot
{R}}{2HR}\right).
\end{equation}
To avoid negative temperature one can also define
$T=|\kappa|/2\pi$. Besides, when $\dot {R}\ll 2HR$, which
physically means that the radius of the apparent horizon is almost
fixed, one may define $T=1/(2\pi R )$ \cite{CaiKim}. We further
assume the energy content of the Universe is in the form of
perfect fluid with energy-momentum tensor
$T_{\mu\nu}=(\rho+p)u_{\mu}u_{\nu}+pg_{\mu\nu},$ where $\rho$ and
$p$ are the energy density and pressure, respectively. As far as
we know, there is no energy exchange between our Universe and out
of its boundary. As a result, we can assume the total
energy-momentum inside the Universe is conserved, which implies
$\nabla_{\mu}T^{\mu\nu}=0$. This leads to
\begin{equation}\label{Cont}
\dot{\rho}+3H(\rho+p)=0.
\end{equation}
In addition, due to the volume change of the Universe, a work
density term is also appeared as \cite{Hay2}
\begin{equation}\label{Work}
W=-\frac{1}{2} T^{\mu\nu}h_{\mu\nu}=\frac{1}{2}(\rho-p).
\end{equation}
Finally, we write down the first law of thermodynamics on the
apparent horizon as
\begin{equation}\label{FL}
dE = T_h dS_h + WdV.
\end{equation}
The total energy inside a 3-sphere with volume $V=4\pi R^3/3$ is
given by  $E= 4\pi \rho R^3/3$. We now calculate the differential
form of the total energy as,
\begin{eqnarray}
\label{dE2}
 dE&=&4\pi R^{2}\rho dR+\frac{4\pi}{3}R^{3}\dot{\rho} dt,\\ \nonumber
 &=&4\pi R^{2}\rho dR-4\pi H R^{3}(\rho+p) dt.
\end{eqnarray}
where in the last step, we have used the continuity equation
(\ref{Cont}).
We propose the entropy associated with the apparent horizon is in
the form of the generalized R\'{e}nyi entropy (\ref{SR2}). In the
cosmological setup, we replace the horizon radius with the
apparent horizon radius $R$. Thus, the explicit form of the
R\'{e}nyi entropy associated with the apparent horizon, up to the
first order correction term, is given by
\begin{equation}\label{SR3}
S_{h}=\mathcal{S}-\frac{\lambda}{2}\mathcal{S}^2,
\end{equation}
where $\mathcal{S}=A/(4G)=\pi R^2/G$ obeys the area law. Taking
differential form of the R\'{e}nyi entropy (\ref{SR3}), we arrive
at
\begin{eqnarray} \label{dSh}
dS_h&=& d \mathcal{S}-\lambda \mathcal{S} d \mathcal{S},
\end{eqnarray}
where
\begin{eqnarray} \label{dS}
d \mathcal{S}= \frac{2\pi R\dot {R}}{G} dt.
\end{eqnarray}
Inserting relations (\ref{Work}), (\ref{dE2}), (\ref{dSh}) and
(\ref{dS}) in the first law of thermodynamics (\ref{FL}) and using
definition (\ref{T}) for the temperature, after some calculations,
we find the differential form of the Friedmann equation as
\begin{equation} \label{Fried1}
\left(1-\lambda\mathcal{S}\right)\frac{dR}{R^{3}}= 4 \pi G
H(\rho+p) dt.
\end{equation}
Using the continuity equation (\ref{Cont}), we reach
\begin{equation} \label{Fried2}
-\frac{2dR}{R^{3}} \left(1-\alpha R^2\right)= \frac{8 \pi G}{3} d
\rho,
\end{equation}
where we have defined $\alpha =\lambda \pi/G$. If we integrate Eq.
(\ref{Fried2}), we immediately reach
\begin{equation} \label{Fried3}
\frac{1}{R^2}-\alpha \ln \left(\frac{1}{R^2}\right)= \frac{8 \pi
G}{3}\rho+\frac{\Lambda}{3},
\end{equation}
where $\Lambda$ is an integration constant which can be
interpreted as the cosmological constant. Substituting $R$ from
Eq.(\ref{radius}), we reach at
\begin{equation} \label{Fried4}
H^2+\frac{k}{a^2}-\alpha \ln \left(H^2+\frac{k}{a^2}\right)=
\frac{8 \pi G}{3}(\rho+\rho_{\Lambda}).
\end{equation}
where $\rho_{\Lambda}=\Lambda/(8\pi G)$.  This is the modified
first Friedmann equation derived when the entropy associated with
the horizon is given by R\'{e}nyi entropy. When $\alpha\rightarrow
0$, we find the Friedmann equations in standard cosmology. For
flat universe ($k=0$), it reduces to
\begin{equation} \label{Fried5}
H^2-\alpha \ln H^2= \frac{8 \pi G}{3}(\rho+\rho_{\Lambda}).
\end{equation}
The second Friedmann equation can be derived by combining Eqs.
(\ref{Cont}) and (\ref{Fried5}). It is a matter of calculation to
show that

\begin{equation} \label{secondFried5}
\dot{H}(1-\alpha H^{-2})=-4\pi G (\rho+p).
\end{equation}
It is important to note that the modified Friedmann equations
derived here, like those in other entropic-gravity approaches
\cite{Sheykhi1,Sheykhi2}, are not obtained from a variational
principle applied to a fundamental gravitational action. Instead,
they follow from the thermodynamics-gravity correspondence applied
to the apparent horizon, using the Clausius relation, the unified
first law, and horizon thermodynamics
\cite{Jac,Pad1,Pad2,CaiKim,Cai2,Sheykhi1}. Within this
phenomenological framework, the continuity equation \eqref{Cont}
is assumed independently, reflecting the condition of no energy
exchange between the horizon and the bulk. Combining Eqs
.(\ref{Cont}), \eqref{Fried4} and \eqref{secondFried5} yields an
identity, ensuring internal consistency analogous to the role of
the Bianchi identities in general relativity. The construction of
a fully covariant action whose FRW reduction reproduces
Eqs.~\eqref{Fried4} and \eqref{secondFried5} remains an
interesting open question for future work.
\section{Constraints on R\'{e}nyi entropy from BBN \label{BBN}}
\subsection{BBN in R\'{e}nyi cosmology}
In this section, we first examine the implications of R\'{e}nyi
cosmology during the radiation-dominated era, followed by an
investigation of BBN in this framework.

Starting from the modified Friedmann equation derived from
R\'{e}nyi entropy given in Eq. \eqref{Fried5}, we consider a flat
universe ($k=0$) and neglect the contribution from cosmological
constant ($\rho_{\Lambda}=0$) at the early stages of the universe.
Therefore Eq. \eqref{Fried5} reads
 \begin{equation} \label{Fr5}
 H^2-\alpha \ln H^2=H_{GR}^2,
 \end{equation}
where $H_{GR}^2=8\pi G\rho  /3$. This equation can also be
rewritten as
 \begin{equation} \label{Fr}
 H^2-\delta=C \;\;\to\; \;H^2=C+\delta,
 \end{equation}
where $C=H_{GR}^2$ and $\delta=\alpha \ln H^2$. Substituting
$H^2=C+\delta$
 in Eq. (\ref{Fr5}), we arrive at
 \begin{equation} \label{Ff1}
    C+\delta-\alpha \ln(C+\delta)=C.
 \end{equation}
Since $\alpha$ is small ($\delta \ll C $), we can use the Taylor
expansion to find
 \begin{equation} \label{Ff2}
    \ln(C+\delta)\simeq \ln C+\frac{\delta}{C}+\mathcal{O}\left( \frac{\delta^2}{C^2}
    \right).
 \end{equation}
 Inserting Eq. \eqref{Ff2} in Eq. \eqref{Ff1} we get
 \begin{multline} \label{Fs}
    \delta-\alpha \left(\ln C+\frac{\delta}{C} \right)=0, \; \;  \Rightarrow  \; \; \delta \left(1-\frac{\alpha}{C} \right)=\alpha \ln C.
 \end{multline}
 Solving for $\delta$ we reach
 \begin{equation} \label{Ffa}
  \delta=\frac{\alpha \ln C}{1-\alpha/C}.
 \end{equation}
 Since $\alpha$ is very small ($\alpha /C \ll1$), we can expand the above
 equation. The result is
 \begin{equation} \label{Ffs}
 \delta\simeq \alpha \ln C\left(1+ \frac{\alpha}{C} \right)\simeq \alpha \ln
 C+\mathcal{O}(\alpha^2).
 \end{equation}
 Replacing $\delta$ in Eq. \eqref{Fr},  to first order in $\alpha$,
 the solution is $H^2\simeq C+\alpha \ln C$. After expanding, the modified Hubble parameter is derived as
 \begin{equation} \label{Ffd}
    H\simeq H_{GR}\left(1+ \alpha\frac{ \ln H_{GR}^2}{2H_{GR}^2}
    \right),
 \end{equation}
where $H>H_{GR}$ for $\alpha>0$, while  $H<H_{GR}$  for
$\alpha<0$.

We can rewrite the modified Hubble parameter (\ref {Ffd}) in the form
\begin{eqnarray}
    H(T)\equiv Z(T) H_{GR}(T).
\end{eqnarray}
Thus, the amplification factor $Z(T)$ can be expressed as
\begin{equation} \label{ZT1}
Z(T)=\frac{H}{H_{GR}(T)} \rightarrow \notag Z(T)=\left(1+
\frac{\alpha \ln H_{GR}^2}{2H_{GR}^2} \right).
\end{equation}
By inserting the relativistic particle energy density, $\rho(T) =
{\pi^2 g_{}T^4}/{30}$, where $g_{} \sim 10$ represents the
effective degrees of freedom and $T$ denotes temperature, along
with the standard GR Hubble parameter $H_{GR}$ and the coupling
parameter $\alpha = {\lambda \pi}/{G}$ into the the above
equation, we derive the amplification factor $Z(T)$ in the
following form
\begin{equation}\label{zt}
Z(T)=1+45\lambda \frac{ \ln
\left(\frac{4\pi^3gGT^4}{45}\right)}{8G^2\pi^2gT^4}.
\end{equation}
As anticipated, for $\lambda=0$, the amplification factor
simplifies to $Z(T)=1$, restoring the GR limit.

\subsection{Primordial light elements{ \textit{ $_{}^{4}\textrm{He}$, D} and \textit {Li} } in R\'{e}nyi cosmology}

Now we examine the permissible range of the R\'{e}nyi parameter
$\lambda$ by studying its cosmological consequences for primordial
nucleosynthesis. Specifically, we investigate how
R\'{e}nyi-modified cosmology affects the predicted abundances of
light elements, particularly Deuterium $_{}^{2}\textit{H}$,
Tritium $_{}^{7}\textit{Li}$, and Helium $_{}^{4}\textit{He}$.
Through systematic comparison between theoretical abundance
predictions and precision observational data, we establish
constraints on the the R\'{e}nyi entropy parameter. The
fundamental innovation of our approach lies in replacing the
conventional $Z$-factor (associated with neutrino degrees of
freedom in standard cosmology \cite {Luciano}) with the modified
amplification factor $Z(T)$ derived in Eq. \eqref{zt}. While the
standard cosmological model assumes $Z=1$, this equality no longer
holds when one considers a modified cosmological model, the
presence of additional relativistic species (such as neutrinos) or
other beyond-standard-model scenarios. In such cases, the
appropriate modification to the $Z$-factor would be represented as
follows \cite{Anish,Luciano,Boran}
\begin{align}
Z_{\nu}=\left [ 1+\frac{7}{43} (N_{\nu}-3)\right ]^{1/2}.
\end{align}
The parameter $N_{\nu}$ quantifies the neutrino species count in
this framework, while the baryon-antibaryon asymmetry parameter
$\eta_{10}$ \cite{Adv,Simha} serves as a key input for our
calculations. To specifically probe R\'{e}nyi cosmology's
influence on BBN, we maintain $N_{\nu}=3$ throughout our analysis.
This choice guarantees that any observed departure of $Z$ from its
standard value ($Z=1$) can be uniquely attributed to modifications
arising from R\'{e}nyi cosmology, rather than being conflated with
potential contributions from additional particle species. The
primordial abundances of nuclei produced during BBN depend on both
the baryon density and the expansion rate of the early universe.
The baryon abundance can be measured by comparing the number
density of baryons (nucleons) to the number density of cosmic
microwave background photons.

Since the coupled nonlinear differential equations governing
primordial nucleosynthesis cannot be solved analytically, an
accurate comparison between theoretical predictions and observed
light element abundances (derived from observational data) can
only be achieved through extensive numerical computation. This
makes the relationship between quantities like $Z$ (which is
proportional to the expansion rate), baryon to photon ratio
parameter $\eta_B$, and the available data for $Y_{Li}$, $Y_{He}$,
and $Y_{D}$ unclear. By plotting the standard model's predicted
abundance contours for ${}^{4}{He}$, $D$, and ${}^{7}{Li}$ as
functions of the expansion rate $Z$ and baryon density parameter
$\eta_B$, and comparing them with numerical fits, we observe that
the light element abundances vary smoothly as functions of $Z$ and
$\eta_B$. Therefore, simple relations exist between the parameters
$Z$, $\eta_B$ and light element abundances.

However, the standard model's predicted primordial abundances are
not linearly related to baryon density. Thus, simple linear fits
can be found for the standard model's predicted abundances as
functions of the parameters $Z$ and $\eta_B$.

For our subsequent calculations, we implement the methodology
developed in \cite{Anish,Sahoo}, briefly outlining its essential
features below:

(i)\ \textit {$_{}^{4}\textrm{He}$ abundance}- The following
observational constraint on the primordial {$_{}^{4}\textit{He}$
abundance is obtained from the numerical best-fit analysis
\cite{Kneller,Annu}
\begin{equation} \label{bestfit}
    Y_p = 0.2485 \pm 0.0006 + 0.0016 \left[ (\eta_{10} - 6) + 100 (Z - 1) \right].
\end{equation}
Here $\eta_{10}$ denotes the baryon density parameter and defined
as
 \cite{Adv,Simha}
\begin{equation} \label {et}
    \eta_{10} \equiv 10^{10}\eta_B \equiv 10^{10} \frac{n_B}  {n_{\gamma}} \simeq 6,
\end{equation}
where the baryon-to-photon ratio is fundamentally defined as
$\eta_B \equiv{n_B}/{n_{\gamma}}$ \cite{Wamp}, serving as a
critical parameter in our analysis. In our framework the
amplication factor $Z$, follows from Eq. (\ref{zt}), governs the
modified nucleosynthesis dynamics. Imposing the condition $Z=1$
reproduces the conventional BBN result for $_{}^{4}\textrm{He}$
abundance, which is almost $(Y_p)|_{\rm GR}=0.2485\pm 0.0006$.
Current astrophysical observations, when analyzed assuming
$\eta_{10}=6$, constrain the helium abundance to $Y_p=0.2449\pm
0.004$ \cite{Brain}. The striking agreement between this empirical
value and our theoretical framework given in Eq. (\ref{bestfit})
reveals
\begin{equation}
    0.2449 \pm 0.0040 = 0.2485 \pm 0.0006 + 0.0016 \left[ 100(Z - 1) \right].
\end{equation}
And hence, we can derive the value of $Z$ as
\begin{equation} \label {zhe}
    Z = 1.0475 \pm 0.105.
\end{equation}

(ii) \ \textit {$_{}^{2}\textit{H}$ abundance}- The primordial
formation of deuterium occurs through the fundamental process:
\begin{equation*}
n + p \rightarrow , ^2\text{H} + \gamma
\end{equation*} Following the methodology of \cite{Adv},
the primordial Deuterium abundance via numerical optimization, can be determined as
\begin{equation} \label {zde}
Y_{D_p} = 2.6(1 \pm 0.06) \left(\frac{6} {\eta_{10} - 6(Z - 1)}
\right)^{1.6}.
\end{equation}
Again setting $Z=1$ and $\eta_{10}=6$ gives the standard GR value
$Y_{D_p}|{GR}=2.6\pm 0.16$. Matching observational data for
Deuterium abundance $Y_{D_p}=2.55\pm 0.03$ \cite{Brain}, with Eq.
\eqref{zde} yields
\begin{equation}
2.55 \pm 0.03 = 2.6(1 \pm 0.06) \left(\frac{6} {\eta_{10} - 6(Z -
1)} \right)^{1.6}.
\end{equation}
We therefore obtain the following constraint on the amplication parameter $Z$
\begin{equation} \label {zobs}
Z = 1.062 \pm 0.444.
\end{equation}
We observe a partial concordance between this constraint and those
derived from the $_{}^{4}\textit{He}$ abundance in \eqref{zhe}.

(iii)\ \textit{$_{}^{7}\textit{Li}$ abundance}- The baryon density
parameter $\eta_{10}$, as defined in Eq. \eqref{et}, provides
excellent agreement with observed primordial abundances of
${}^{4}\textit{He}$ and $D$, yet reveals the long-standing Lithium
discrepancy in BBN. While standard BBN calculations with
$\eta_{10} \approx 6$ successfully match light element abundances
for deuterium ($Y_D$) and helium-4 ($Y_p$), they systematically
overpredict the $_{}^{7}\textit{Li}$ abundance by a factor of
$3-4$ compared to observational data. Comparison reveals that the
predicted $_{}^{7}\textit{Li}$ abundance in $\Lambda$CDM cosmology
systematically overestimates observational values, with the exact
discrepancy range shown in \cite{Boran}
\begin{align*}
    \frac{\textit{Li}|_{\rm GR}}
    {\textit{Li}|_{\rm obs}}
    \in [2.4 - 4.3].
\end{align*}
Surprisingly, the standard BBN framework, while remarkably
successful in predicting the abundances of ${}^{4}\textit{He}$ and
$D$, fails to account for the observed ${}^{7}\textit{Li}$
abundance in metal-poor halo stars. This unresolved tension, known
as the \textit{Cosmological Lithium Problem} \cite{Boran},
presents a significant challenge to our understanding of
early-universe physics. The numerical best fit for the abundance
of $_{}^{7}\textit{Li}$ is given by
\begin{equation}\label{dd}
    Y_{Li} = 4.82 (1 \pm 0.1)\left[\frac{\eta_{10} - 3(Z - 1)} {6} \right]^{2},
\end{equation}
By requiring consistency between:
\begin{itemize}
\item The observational Lithium abundance $Y_{Li} = 1.6 \pm 0.3$
\cite{Brain}.
\item The numerical best-fit prediction for $_{}^{7}\textit{Li}$ abundance provided in Eq.
\eqref{dd}.
\end{itemize}
The modified expansion-rate parameter $Z$ is constrained to the
interval
\begin{equation} \label {lit}
Z = 1.960025 \pm 0.076675.
\end{equation}
The derived bounds conflict with both ${}^{4}\textit{He}$ (Eq.
\eqref{zhe}) and ${}^{2}\textit{H}$ (Eq. \eqref{zobs}) abundance
measurements, revealing the fundamental Lithium problem.
\subsection{Discussion on the Implications of R\'{e}nyi Cosmology to Lithium Problem}
Here, building upon our previous analysis, we consolidate the
principal findings and analyze the results. Our analysis
constrains the R\'{e}nyi parameter $\lambda$ using observational
bounds from primordial nucleosynthesis. Fig. 1 illustrates the
relation in Eq. \eqref{zt}, incorporating constraints derived from
Eq. \eqref{zhe}. As we can see, the $_{}^{4}\textit{He}$ abundance
measurements limit $\lambda$ to
\begin{equation} \label {dhes}
-3.18 \times10^{-85}\lesssim \lambda \lesssim 1.19 \times10^{-85},
\end{equation}
where we have used the reduced Planck mass $M_p = (8\pi G)^{-1/2}
\simeq 2.4 \times 10^{18}$ GeV in our numerical calculations
\cite{Luciv}.
\begin{figure} [H]
 \includegraphics[scale=0.88]{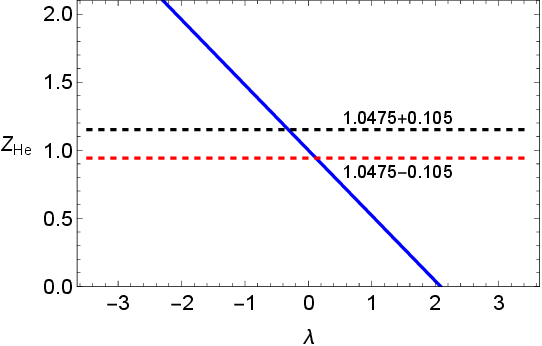}
\caption{$Z_{He}$ vs R\'{e}nyi parameter $\lambda $. The
observational interval is reported in (\ref {zhe}). We have set
$\eta_{10}=6$ and the freeze-out temperature $T_f=10 MeV$.}
 \label{Fig1}
\end{figure}
Fig. 2 displays the results for primordial deuterium ($D$)
abundance using the relation in Eq. \eqref{zobs}. The analysis
reveals the following constraint on the R\'{e}nyi parameter
\begin{equation} \label {deuts}
    -1.054 \times 10^{-85}\lesssim \lambda \lesssim  7.96 \times 10^{-85}.
\end{equation}

\begin{figure} [H]
    \includegraphics[scale=0.86]{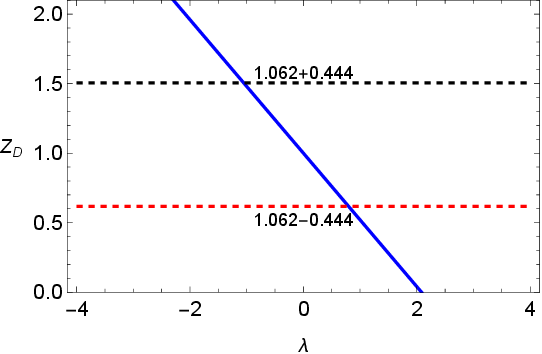}
\caption{$Z_D$ vs R\'{e}nyi parameter $\lambda $. The
observational interval is reported in (\ref {zobs}). We have set
$\eta_{10}=6$ and the freeze-out temperature $T_f=10 MeV$.}
    \label{Fig2}
\end{figure}
The permitted range for the R\'{e}nyi parameter $\lambda$ derived
from deuterium analysis (Eq. \ref{deuts}) shows agreement with the
constraints obtained from helium-4 measurements (Eq. \ref{dhes}).
The results for the Lithium are shown in Fig. 3. The plot presents
the constraints on the R\'{e}nyi parameter $\lambda$ derived from
primordial $_{}^{7}\textit{Li}$ abundance analysis as
\begin{equation} \label {ZLi}
-2.16 \times^{-84}\lesssim \lambda \lesssim -1.84 \times 10^{-84}.
\end{equation}
As we can see from Fig. 3, the analysis of primordial
${}^{7}\textit{Li}$ abundance indicates the permitted range of the
R\'{e}nyi parameter $\lambda$ exhibits no consistency with the
constraints obtained from ${}^{4}\textit{He}$ (Eq. \ref{dhes}) and
$_{}^{2}\textit{H}$ (Eq. \ref{deuts}) abundances.\\
Note that the analysis presented here employs the linear expansion
of the Rényi entropy. For the observationally allowed range of
$\lambda$ derived from BBN ($\lambda \lesssim 10^{-85}$ in our
units), the dimensionless combination $\lambda\mathcal{S}$ during
the BBN epoch satisfies $\lambda\mathcal{S} \ll 1$. Consequently,
including the next-order term ${\lambda^2}\mathcal{S}^3/{3}$ from
the full expansion \eqref{SR2} would modify the Hubble parameter
$H(T)$ and the amplification factor $Z(T)$ by a fraction less than
$ 1\%$, well below current observational uncertainties. The
truncation to linear order in $\lambda$ is therefore fully
justified within the phenomenologically relevant parameter space.
Values of $\lambda$ large enough to make higher-order terms
important are excluded by BBN data.

\begin{figure} [H]
    \includegraphics[scale=0.88]{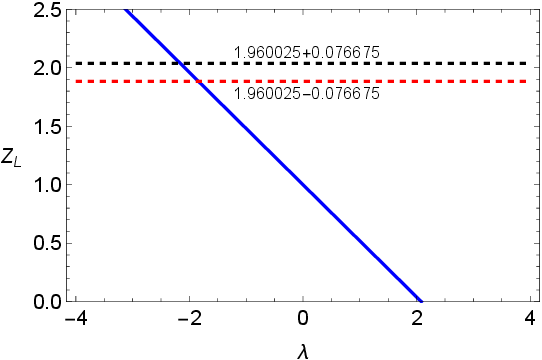}
\caption{$Z_{Li}$ vs R\'{e}nyi parameter $\lambda $. The
observational interval is reported in (\ref {lit}). We have set
$\eta_{10}=6$ and the freeze-out temperature $T_f=10 MeV$.}
    \label{Fig3}
\end{figure}
The \textit{Lithium problem} stems from the mismatch between
theoretical predictions and observed amounts of primordial
$_{}^{7}\textit{Li}$ from BBN. If a modified cosmological model's
free parameter can successfully explain the measured abundances of
$_{}^{4}\textit{He}$ and deuterium (which agree with standard BBN)
and the problematic Lithium observations, then we could
potentially resolve the Lithium discrepancy without needing
astrophysical explanations.

Current constraints show that the R\'{e}nyi parameter $\lambda$
fits both helium and deuterium data well, however, no valid
$\lambda$ values can simultaneously account for the Lithium
measurements. Therefore, no common parameter range exists for the
R\'{e}nyi parameter $\lambda$ that can simultaneously explain the
observed abundances of all three light elements
${}^{4}\textit{He}$, $D$, and ${}^{7}\textit{Li}$ at the same
time, and hence, resolve the Lithium discrepancy. However, the
permitted $\lambda$ ranges for $_{}^{4}\textit{He}$ and $D$ show
only minor deviations from the range obtained for
$_{}^{7}\textit{Li}$. This opens a new window on the resolving the
Lithium problem within modified cosmological frameworks.

It should be noted that the derived parameter ranges in Eqs.
\eqref{dhes}, \eqref {deuts} and \eqref{ZLi} are obtained for a
specific temperature of $T=10MeV$ and will vary at different
energy scales. However, the key findings remain, it means across
all relevant temperatures in the early universe, the allowed
$\lambda$ range for $_{}^{7}\textit{Li}$ shows no overlap with
$_{}^{4}\textit{He}$ and $D$ constraints, and they have minimal
deviation from $\Lambda CDM$ predictions.\\

The R\'{e}nyi entropy framework studied here can be compared with
other entropic generalizations that modify early-universe
dynamics. Tsallis $\beta$-entropy introduces power-law corrections
to the area law, typically constrained to $\beta \sim 10^{-1}$ by
BBN \cite{Anish, Jizba}. Barrow entropy corresponds to a
fractal-dimension deformation $A^{1+\delta/2}$ with bounds $\delta
\sim 10^{-2}$ \cite{Ava1, Barrow}. Kaniadakis statistics yields
relativistic-like deformations, limited to $\tilde{K} \sim
10^{-78}$ \cite{Ava2}. In contrast, the R\'{e}nyi logarithmic
correction leads to a distinct functional form in the Friedmann
equations, and our BBN analysis constrains the R\'{e}nyi parameter
to $\lambda \sim 10^{-85}$, which is orders of magnitude smaller.
This indicates that the R\'{e}nyi deformation is
phenomenologically much milder during BBN, producing negligible
deviations from standard cosmology for the allowed parameter
range, while still sharing the persistent tension between D/$^4$He
and $^7$Li predictions common to many modified entropic scenarios.

\subsection{Outlook: Towards Resolving the Lithium Problem} \label{subsec:outlook}
While the simple constant-$\lambda$ R\'{e}nyi model explored in
this work does not simultaneously reproduce the observed
abundances of D, $^4$He, and $^7$Li, the proximity of the
$\lambda$ ranges allowed by each element suggests that addressing
the Lithium problem within an extended R\'{e}nyi framework remains
feasible. This possibility arises from the distinct
temperature-dependent chronology of BBN. Neutron-proton freeze-out
occurs at $T \sim 0.8$ MeV, setting the initial neutron-to-proton
ratio. Deuterium and helium-4 form at $T \sim 0.1$ MeV, when
nuclear reaction rates are fast compared to the Hubble expansion,
rendering their abundances largely insensitive to modest changes
in the expansion rate. In contrast, $^7$Li production takes place
at lower temperatures, $T \sim 0.03$--$0.06$ MeV, through slower
secondary reactions that are highly sensitive to the expansion
rate and the available reaction time. Consequently, modifications
to the Hubble rate, such as those induced by generalized
entropies, can selectively alter the predicted $^7$Li abundance
while leaving D and $^4$He essentially unchanged. In the R\'{e}nyi
framework, such modifications enter through corrections to the
horizon entropy-area relation, which translate into new terms in
the Friedmann equations that can be interpreted either as a
modified expansion rate or as an effective energy density.

Building on the constraints derived here, several concrete
extensions could be explored in future work. A
temperature-dependent parameterization $\lambda(T)$ would allow
the correction to affect primarily the later $^7$Li production
epoch. Hybrid entropic-scalar models that combine R\'{e}nyi
entropy with additional light fields or varying constants could
provide more flexibility to fit all light-element data.
Multi-parameter entropic generalizations that interpolate between
different nonextensive forms might capture more nuanced
thermodynamic signatures of quantum spacetime.

The present analysis therefore provides a necessary
phenomenological baseline: it demonstrates that the
constant-$\lambda$ R\'{e}nyi model is ruled out as a complete
solution, but it also shows that entropy-induced modifications can
meaningfully shift the predicted $^7$Li abundance. This insight
offers a clear theoretical pathway for constructing more
realistic, observationally consistent entropic cosmologies while
preserving the remarkable agreement of standard BBN with D and
$^4$He measurements.

\section{Implication of R\'{e}nyi Cosmology on baryogenesis \label{br}}
Our aim in this section is to check the consistency with
observational data on baryogenesis within the framework of
R\'{e}nyi cosmology and constrain the R\'{e}nyi parameter as well.
In \cite{Lym}, the additional terms dependent on Kaniadakis
parameter $K$ that appear in the modified Friedmann equations are
interpreted as contributions to effective dark energy density and
pressure. In this discussion, we adopt a broader perspective and
assert that R\'{e}nyi entropy introduces fluctuations in energy
density and pressure relative to the BG-like equilibrium.
Consequently, we reformulate the total energy density and pressure
to incorporate these adjustments as follows

\begin{align}
 \rho &= \rho_0 + \delta\rho_R\,, \label{eq:18} \\
 p &= p_0 + \delta p_R\,. \label{eq:19}
\end{align}
Substituting Eqs.~\eqref{eq:18} and \eqref{eq:19} into the
modified Friedmann equations, we derive the leading order results
as

\begin{eqnarray}\label{dr}
&& H^2-\alpha \ln H^2=\frac{8\pi G}{3}(\rho_{0}
+\delta \rho _{R}), \nonumber \\
&&\Rightarrow \frac{8\pi G}{3}\rho_{0}-\alpha \ln\left(\frac{8\pi
G}{3}\rho_{0}\right)=\frac{8\pi G}{3}(\rho_{0} +\delta \rho _{R}),\nonumber \\
&& \Rightarrow \delta \rho_{R}=-\frac{3\alpha}{8\pi
G}\ln\left(\frac{8\pi G}{3}\rho_{0}\right).
\end{eqnarray}
Inserting $\alpha=\lambda \pi /G$ into Eq. \eqref{dr} we have
\begin{equation}\label{rro}
\delta \rho_{R}=-\frac{3\lambda}{8G^2} \ln\left(\frac{8\pi
G}{3}\rho_{0}\right).
\end{equation}
Following the same approach, if we insert Eqs.~\eqref{eq:18} and
\eqref{eq:19} into the second friedmann equation in \eqref
{secondFried5} and combining the result with \eqref{rro}, for the
radiation dominated era ($\omega=1/3$), we obtain
 \begin{equation}\label{prr}
 \delta p_{R}=\frac{\lambda}{8G^2} \left[3\ln\left(\frac{8\pi
 G}{3}\rho_{0}\right)-4\right].
 \end{equation}
In this context, we have incorporated all the $\lambda$-dependent
terms into $\delta\rho_R$ and $\delta p_R$. As expected, both of
these quantities approach zero as $\lambda\rightarrow 0$
approaches zero, which aligns with the reestablishment of the
standard cosmological model in this limit.

We will now explore how the fluctuations in mass density and
pressure, as described in equations \eqref{rro} and \eqref{prr},
influence cosmic evolution, particularly the process of
baryogenesis. Observational evidence suggests that matter is more
abundant than antimatter in our Universe, which sharply contrasts
with the predictions made by Quantum and Relativistic theories, as
discussed in detail in \cite{ref41}. The conventional cosmological
model proposes that baryogenesis occurs dynamically as the
Universe expands and cools. Specifically, \cite{ref42} outlines
three essential conditions for this mechanism to take place: (i) a
violation of baryon number B, which is necessary to create an
excess of baryons over antibaryons; (ii) violations of C-symmetry
and CP-symmetry. The first condition ensures that interactions
leading to a greater production of baryons than antibaryons are
not balanced out. Similarly, CP-violation is crucial to prevent
the equal production of left-handed baryons and right-handed
antibaryons (and vice versa); (iii) interactions that are out of
thermal equilibrium, as otherwise, CPT symmetry would balance the
processes that increase and decrease the baryon number.

In our framework, the first two Sakharov conditions are fulfilled
by incorporating the standard interaction between baryon current
and spacetime. Conversely, the final condition is achieved by
disrupting thermal equilibrium through the modified Friedmann
equations discussed in the previous section. Additionally, it is
worth noting that a similar investigation has been carried out in
\cite{ref43}, focusing on gravitational modifications of the
Heisenberg Uncertainty Principle. In that scenario, the mechanism
for the deviation from equilibrium is attributed to minimal length
effects stemming from a phenomenologically motivated quantum
gravity framework. Further analysis in this area is also presented
in \cite{ref44}, which explores $f(T)$ gravitational baryogenesis.

In supergravity theories, as noted in \cite{ref43}, a mechanism
has been proposed for generating baryon asymmetry through a
dynamical breaking of $CPT$ (and $CP$) during the Universe's
expansion \cite{ref45}. Although this mechanism satisfies the
first two Sakharov conditions, it maintains thermal equilibrium,
thereby contravening the Sakharov protocol. Within this framework,
the interaction that leads to $CPT$ violation is characterized by
a coupling between the derivative of the Ricci scalar $\mathcal{R}$ and the
baryon current $J^\mu$, as described in \cite{ref46}.
 \begin{equation}
 \frac{1}{M^2_*} \int d^4 x \sqrt{-g}\, J^\mu \partial_\mu \mathcal{R}. \label{eq:22}
 \end{equation}
The determinant of the metric tensor is denoted by $g$, and the
cutoff scale of the effective theory is represented by $M_*$,
which is approximately equal to the reduced Planck mass, given by
$M_* = (8\pi G)^{-1/2} \approx 2.4 \times 10^{18}$ GeV.

By imposing an interaction that violates baryon number $B$ while
maintaining thermal equilibrium, in accordance with the first
Sakharov condition, a net asymmetry between matter and antimatter
can be generated and subsequently frozen below the decoupling
temperature $T_D$, at which point $B$-violation ceases to be in
equilibrium. Specifically, this is illustrated by
Eq.~\eqref{eq:22} as referenced in \cite{ref46}. We find
\begin{equation}
\frac{1}{M^2_*} J^\mu \partial_\mu \mathcal{R} = \frac{1}{M^2_*} (n_B -
n_{\bar{B}}) \dot{\mathcal{R}}. \label{eq:23}
\end{equation}
Here the baryon number density is represented by $n_B$, while the
anti-baryon number density is denoted as $n_{\bar{B}}$.

As mentioned in \cite{ref46}, the dynamical violation of $CPT$
alters thermal equilibrium in a manner akin to that of a chemical
potential. According to Eq.~\eqref{eq:23}, the effective potential
expressions for baryons and anti-baryons are given by $\mu_B =
-\mu_{\bar{B}} = -\dot{\mathcal{R}}/M^2_*$. Consequently, the net baryon
number density in the early Universe can be expressed as follows
 \begin{equation}
 n_B - n_{\bar{B}} = \left|\frac{g_b}{6} \mu_B T^2\right|. \label{eq:24}
 \end{equation}
The parameter $g_b$ is approximately of order one, representing
the number of degrees of freedom associated with baryons.

Baryon asymmetry is currently expressed through the introduction
of a specific parameter \cite{ref47}.
\begin{equation}\label{eq:25}
\frac{\eta} {7}\equiv \frac{n_B - n_{\bar{B}}}{s} = \left|\frac{15
g_b}{4\pi^2 g_{*s}} \frac{\dot{\mathcal{R}}}{M^2_* T}\right|.
\end{equation}
The right-hand side must be assessed for $T = T_D$. Here, we
define $s = 2\pi^2 g_{*s} T^3/45$ as the entropy density during
the radiation-dominated era, where $g_{*s}$ represents the number
of degrees of freedom for particles that contribute to the
Universe's entropy. As noted in \cite{ref47}, the approximation
$g_{*s} \approx g_*$ is valid for this analysis, with $g_* \approx
106$ being the total number of degrees of freedom for relativistic
Standard Model particles in the context of baryogenesis.

From Eq.~\eqref{eq:25}, it can be concluded that $\eta \neq 0$ if
the Ricci scalar curvature changes over time. In standard
cosmology, which is based on black hole entropy, $\dot{\mathcal{R}}=0$,
during the radiation-dominated era, as thermal equilibrium is
maintained. This leads to $\eta = 0$, resulting in equal
production of baryons and anti-baryons. Conversely, deviations
from equilibrium in R\'{e}nyi Cosmology are likely caused by the
variations in mass density and pressure as described in equations
\eqref{eq:18} and \eqref{eq:19}.

Now, the R\'{e}nyi-corrected derivative of the Ricci scalar
$\dot{R}$ can be determined by noting that
\begin{equation}
\mathcal{R}_{Reny} = -8\pi G T_g\,, \label{eq:26}
\end{equation}
where $T_g = \rho - 3p$ is the trace of the energy--momentum
tensor. Combining Eqs.~\eqref{eq:18} and \eqref{eq:19} with $T_g$,
and using Eqs.~\eqref{rro} and \eqref{prr} we are led to
\begin{align}\label{Tg1}
T_g=(1-3\omega)\rho_{0}-\frac{3\lambda}{2G^2}\left[\ln\left(\frac{8\pi
G}{3}\rho_{0}\right)-1\right].
 \end{align}
Inserting the above equation in Eq.~\eqref{eq:26} we obtain
\begin{equation} \label{eq:27}
\mathcal{R}_{Reny} =\mathcal{R}_0 +\lambda
\frac{12\pi}{G}\left[\ln\left(\frac{8\pi
G}{3}\rho_{0}\right)-1\right]\,,
 \end{equation}
where we have defined the equilibrium Ricci scalar $\mathcal{R}_0$ as
\begin{equation}
\mathcal{R}_0 = -8\pi G (1 - 3w) \rho_0\,. \label{eq:28}
\end{equation}
For the radiation dominated era ($w = 1/3$), however, one simply
has $\mathcal{R}_0 = 0$, giving
\begin{equation}
\mathcal{R}_{Reny} = \lambda
\frac{12\pi}{G}\left[\ln\left(\frac{8\pi
G}{3}\rho_{0}\right)-1\right]\,. \label{eq:29}
\end{equation}
Now, the time derivative of the Ricci scalar takes the form
\begin{equation}
\dot{\mathcal{R}}_{Reny} =- 16\sqrt{8\pi^3} \lambda
\;\sqrt{\frac{3\rho_0}{G} } \neq 0\,, \label{eq:30}
\end{equation}
where we have used the continuity equation  for the radiation
dominated era $\dot{\rho}_{0} =-4H\rho_0$.

From the preceding equation, it is evident that employing
R\'{e}nyi entropy to account for the horizon degrees of freedom in
the Universe offers a coherent framework for describing deviations
from thermal equilibrium, thereby fulfilling the third and final
Sakharov condition. Notably, as $\lambda$ approaches zero, the
conventional cosmological model with $\dot{\mathcal{R}} = 0$ is reinstated.

Next, we substitute Eq.~\eqref{eq:30} into the baryon asymmetry
equation \eqref{eq:25}. A simple calculation yields
\begin{equation}
\eta=420\lambda\;\sqrt{\frac{8}{\pi}}\frac{g_b
}{g_{*s}M_*^2T_D}\sqrt{\frac{3\rho_0}{G}}\,. \label{eq:32}
\end{equation}
This can be further refined by expressing the equilibrium mass
density $\rho_0$ as a function of temperature, given by $\rho(T) =
{\pi^2 g_{}T^4}/{30}$. Substituting this expression into
Eq.~\eqref{eq:32} results in

\begin{equation}
\eta=840 \lambda\sqrt{\frac{\pi}{5}}\;\frac{T_D
}{M_*^2\sqrt{g_{*}G}}, \label{eq:33}
\end{equation}
where we have used $g_b \sim \mathcal{O}(1)$ and $g_{*s} \approx
g_*$, as discussed above. Following \cite{ref43}, we finally set
$T_D = M_I$, where $M_I \approx 3.3 \times 10^{16}$ GeV is the
upper bound on the tensor mode fluctuation constraints in the
inflationary scale \cite{ref46}.

In order to constrain the R\'{e}nyi parameter, let us consider
observational bounds on baryon asymmetry. From
\cite{ref48,ref49,ref50,ref51,ref52}, it is known that current
measurements give
\begin{equation}
\eta \lesssim 9.9 \times 10^{-11}\,, \label{eq:35}
\end{equation}
 which constrains the $\lambda$-parameter in units of $k_B$ to be (for
 $\lambda \geq 0$, see Fig. 4)
 \begin{equation}
  \tilde{\lambda} \lesssim 0.022\,. \label{eq:36}
 \end{equation}
 \begin{figure} [H]
 \includegraphics[scale=0.86]{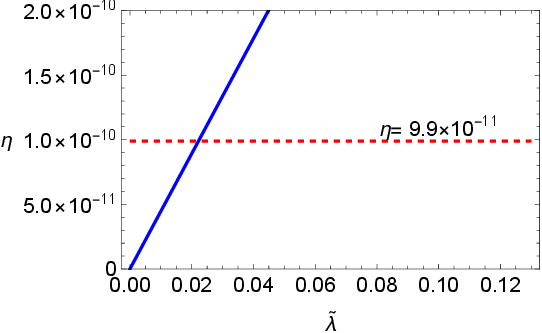}
 \caption{ Baryon asymmetry parameter $\eta$ vs re-scaled R\'{e}nyi parameter $\tilde{\lambda} $.}
 \label{Fig4}
 \end{figure}
We have defined the re-scaled R\'{e}nyi parameter as
$\tilde{\lambda} = \lambda/10^{-9}$. This finding reinforces the
approximation of $\lambda\ll 1$ that we have utilized throughout
the preceding analysis. By comparison between Eq.~\eqref{eq:36}
and the bounds on $\lambda$ in Eqs.~ \eqref{dhes}, \eqref{deuts}
and \eqref{ZLi} it is worth noting that there is only partial
overlap between the ranges of $\lambda$.

\section{Time-temperature relation in R\'{e}nyi cosmology \label{tT}}
In this section, we investigate how R\'{e}nyi cosmology modifies
the fundamental relationship between cosmic time and temperature
during the early Universe. The modified Friedmann equations in
this scenario alter the early Universe's thermodynamics, resulting
in a distinct time-temperature evolution compared to standard
cosmology. During the radiation-dominated epoch, the early
universe maintains thermal equilibrium, ensuring entropy
conservation within a comoving volume. This fundamental principle
leads to \cite{Weinberg}
\begin{align}\label{ent1}
s(T)a^3=\rm {const},
\end{align}
where $s(T)$ is the entropy per unit comoving volume. The entropy
conservation equation for a comoving volume in the early universe
leads to an important relation between temperature and time.
Taking the time derivative of Eq. \eqref{ent1}, we obtain
\begin{align}\label{ent2}
\dot{s}(T)a^3+3\dot{a}a^2s(T)=0 \mapsto
\frac{ds(T)}{dt}a^3=-3\dot{a}a^2s(T) \;.
\end{align}
Substituting $H$ from Eq. (\ref{Ffd}) into Eq. (\ref{ent2}), we
derive
\begin{align}\label{ent3}
\frac{ds(T)}{dt}=-3\left( H_{GR}+ \frac{\alpha \ln
H_{GR}^2}{2H_{GR}} \right)s(T).
\end{align}
The above equation can be rewritten in the following form
\begin{align}\label{dtT}
dt=-\frac{ds(T)}{3s(T)}\left( H_{GR}+ \frac{\alpha \ln
H_{GR}^2}{2H_{GR}} \right)^{-1}.
\end{align}
The cosmic time $t$ as a function of temperature $T$ is obtained
by integrating the above equation. We find
\begin{align}\label{dis}
t=-\frac{1}{3}\int_{}^{} \frac{{s}'(T)}{s(T)}\left( H_{GR}+
\frac{\alpha \ln H_{GR}^2}{2H_{GR}} \right)^{-1} dT,
\end{align}
where the prime symbol denotes differentiation with respect to
temperature, such that for any function $f(T)$, we have $f' \equiv
df/dT$. In cosmological epochs dominated by radiation (where the
equation of state $p=\rho/3$ holds), the  entropy density and
energy density follow the relations \cite{Weinberg}
\begin{eqnarray}\label{3ss}
&&s(T)=\frac{2\mathcal{N}a_B T^3}{3},\\ &&
\rho(T)=\frac{\mathcal{N}a_B T^4}{2}.\label{ro}
\end{eqnarray}
For a relativistic gas $\mathcal{N}$ enumerates all particles and
antiparticles degrees of freedom, counting distinct spin states
separately \cite{Weinberg}. Through algebraic expansion and
insertion of the relation  $\frac{{s}'(T)}{s(T)}=\frac{3}{T}$ ino
Eq. \eqref{dis}, we obtain
\begin{align}\label{above}
t=-\int_{}^{}\frac{1}{TH_{GR}}\left( 1-\alpha\frac{ \ln
H_{GR}}{H_{GR}^2} \right)dT.
\end{align}
Substitution of the aforementioned $\rho(T)$ relation into Eq.
\eqref{above}, yields
\begin{align}\label{tR}
t=  -\int_{}^{}\left( \frac{1}{\beta T^3} -\alpha \frac{\ln (\beta
T^2)}{\beta^{3}T^{7}}\right)dT,
\end{align}
where we have defined $\beta \equiv \sqrt{\frac{8\pi G\mathcal {N}
a_{B}}{6c^2}} $. We first calculate the second term in Eq.
(\ref{tR}) which is
\begin{align}\label{Ist}
I_{st}=\int_{}^{}\frac{ \alpha \ln (\beta T^2)}{\beta^{3}T^{7}}dT
=\frac{\alpha}{\beta^3}\int_{}^{}\left(\frac{\ln
\beta}{T^7}+\frac{2\ln T}{T^7}\right)dT.
\end{align}
Integrating by parts, yields
\begin{align}\label{Int2}
I_{st}=-\frac{\alpha}{6\beta^3T^6}\left(\ln (\beta
T^2)+\frac{1}{3}\right)+ \rm const.
\end{align}
Therefore we find the time-temperature relation in the framework
of R\'{e}nyi Cosmology  as
\begin{align}\label{tTRe}
t=\frac{1}{2\beta T^2}-\frac{\alpha}{6\beta^3T^6}\left(\ln (\beta
T^2)+\frac{1}{3}\right) +\rm const.
\end{align}
In the limiting case where $\alpha = 0$, the time-temperature
relation reduces to $t=\frac{1}{T^2}\sqrt{\frac{3c^2}{16\pi
G\mathcal{N}a_{B}}} $ which exactly matches the standard GR result
\cite{Weinberg}. Note that while our modified Friedmann equations
were derived in natural units ($\hbar = k_B = c = 1$), dimensional
consistency requires that the right-hand side of Eq. \eqref{tTRe}
must have dimensions of time $[t]$, and hence all fundamental
constants must be properly restored. When including all physical
constants explicitly, the parameter $\alpha$ takes the form
\begin{equation}
\alpha\equiv \frac{\lambda \pi }{Gc^2}\mu^2 \gamma,  \quad \
\mu\equiv {k_B c^3\over \hbar},\quad \quad \gamma\equiv {\hbar
c\over k_B}.
\end{equation}
Therefore, the right-hand side of Eq. \eqref{tTRe} naturally
possesses the correct time dimension when all fundamental
constants are properly invoked.

In the early universe's hot plasma, the relativistic particles
content included: photons, three generations of neutrinos,
antineutrinos, and electron-positron pairs giving the total
effective degrees of freedom as $\mathcal{N} ={43}/{4}$. Therefore
in cgs (centimeter-gram-second) units, Eq. \eqref{tTRe} gives,
\begin{equation} \label{tTr}
t=0.994\;\left( {T\over10^{10} K^{\circ}}
\right)^{-2}F(\lambda,T)+\rm const.,
\end{equation}
where $ F(\lambda,T)$ is defined as
\begin{equation}
F(\lambda,T)\equiv 1- \frac{1.98
\times10^{104}}{T^4}\lambda\left[\frac{1}{3}+\ln(5.019\times
10^{-21}T^2)\right].
\end{equation}
The $\lambda \to 0$ limit eliminates all modifications ($F \to
1$), guaranteeing consistency with standard GR \cite{Weinberg}.

Fig. 5 displays the evolution of temperature $T$ versus cosmic
time $t$ across various values of the R\'{e}nyi parameter
$\lambda$, demonstrating that larger values of R\'{e}nyi parameter
causes higher temperatures in the early universe.
\begin{figure} [H]\label{Fig5}
\includegraphics[scale=0.9]{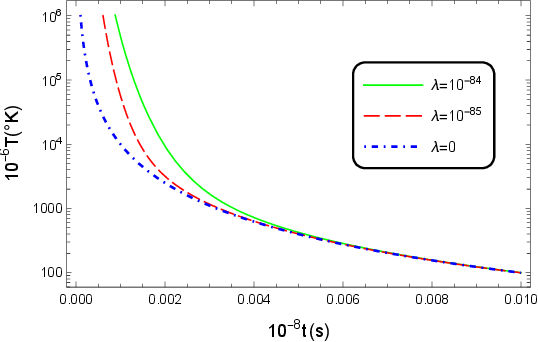}
\caption{The behavior of the temperature vs time in the radiation
dominated era of R\'{e}nyi cosmology for different values of the
R\'{e}nyi parameter $\lambda$.}
\end{figure}
The temperature dependence on the R\'{e}nyi parameter $\lambda$
arises fundamentally from nonextensive thermodynamics in this
modified cosmological framework. Unlike standard cosmology where
thermodynamics is extensive, R\'{e}nyi entropy introduces
non-additive effects that enhance particle interactions and
correlations. These nonextensive properties effectively increase
the energy density compared to GR, necessitating higher
temperatures to maintain equivalent energy densities. Furthermore,
the modified Friedmann equations in this theory alter the early
universe's expansion dynamics - larger $\lambda$ values reduce the
expansion rate during the earliest epochs. This slower expansion
prolongs high-temperature conditions by extending thermalization
timescales. Such behavior aligns with broader nonextensive
thermodynamic theories like Tsallis cosmology, where similar
deformation parameters produce analogous high-energy modifications
to standard thermodynamic relations. The combined effects of these
nonextensive thermodynamic properties and altered expansion
dynamics explain the temperature enhancement observed with
increasing $\lambda$ values.
\section{Closing remarks } \label{Con}
Primordial BBN is one of the three strong evidences for the
Big-Bang model of cosmology together with the expansion of the
Universe and the Cosmic Microwave Background radiation. On the
other hand, the so called cosmological\textit{ Lithium problem},
remains unresolved in the context of standard cosmology. This
motivates people to either search for a new physics beyond the
standard BBN framework or modify the cosmological field equations.

In this work we adopt the viewpoint of modification of
cosmological field equations through thermodynamics-gravity
conjecture. In this way, any modification to the entropy of the
apparent horizon, leads to the modified cosmological field
equations. We have taken into account the R\'{e}nyi entropy, a
nonextensive generalization of Boltzmann-Gibbs entropy. We
disclosed that the origin of MOND theory can be completely
understood through modification of Newton's law of gravitation. In
the relativistic regime, and using the first law of
thermodynamics, this leads to a correction term in the Friedmann
equations.

We then derived stringent constraints on the R\'{e}nyi entropy
parameter by comparing its predicted effects on BBN with precision
observations of light-elements abundances and baryogenesis. By
analyzing BBN and baryogenesis observations, we have established
bounds on the R\'{e}nyi entropy parameter $\lambda$. This
parameter introduces modifications to the Friedmann equations
through nonextensive thermodynamic corrections, which must remain
sufficiently small to maintain agreement with primordial abundance
measurements. We observed that, deviation from the standard
cosmological model is small, as expected. Additionally our
analysis of primordial light element abundances --
${}^{4}\textit{He}$, ${}^{2}\textit{H}$, and ${}^{7}\textit{Li}$
-- reveals that while the allowed $\lambda$ ranges for
${}^{4}\textit{He}$ and ${}^{2}\textit{H}$ show significant
overlap, they exhibit a slight tension with constraints from
${}^{7}\textit{Li}$ abundances. The Lithium problem involves a
mismatch between the predicted and observed primordial Lithium
abundances derived from BBN. If there exists a range of values for
the free parameter in a modified cosmological model that can
simultaneously match the observed abundances of helium-4
($_{}^{4}\text{He}$) and deuterium (which align with standard BBN
predictions), as well as lithium, then it might be possible to
resolve the Lithium discrepancy without relying on astrophysical
adjustments. This minor inconsistency between the overlapping
range of $\lambda$  from $_{}^{4}\text{He}$ and deuterium
abundances and the allowed range obtained from lithium-7 indicates
that addressing the Lithium problem within the context of altered
cosmological frameworks could be feasible. We have also
demonstrated that the modifications introduced by R\'{e}nyi
entropy to the mass density and pressure naturally induce a
mechanism capable of pushing the early Universe out of thermal
equilibrium. This satisfies the third Sakharov condition for
baryogenesis. When combined with a space-time-baryon current
coupling that meets the first two Sakharov requirements, this
framework provides all the necessary elements for successful
baryogenesis. By ensuring agreement between theoretical
predictions and observational constraints, we have derived the
bound \eqref{eq:36} on the R\'{e}nyi parameter. By comparison
between the range of $\lambda$ derived from baryogenesis in
Eq.~\eqref{eq:36} (which is small as expected) and the bounds on
$\lambda$ from light element abundances in Eqs.~\eqref{dhes},
\eqref{deuts} and \eqref{ZLi} it is worth noting that there is a
partial overlap between the ranges of $\lambda$. Furthermore, we
establish the cosmic time-temperature relation $t(T)$ in this
framework, demonstrating that increasing $\lambda$ elevates
early-universe temperatures. This thermal enhancement arises from
two key effects: (i) nonextensive thermodynamics modifies the
energy density through enhanced particle interactions, and (ii)
the modified expansion rate prolongs high-temperature conditions.
These results align with those of other nonextensive gravitational
theories, such as Tsallis cosmology, where similar deformation
parameters produce comparable high energy modifications.
\acknowledgments{We thank Shiraz University Research Council. We
thank referees for very constructive comments which helped us
improve our paper significantly. This work is based upon research
funded by Iran National Science Foundation (INSF) under project
No. 4022705.}

\end{document}